\documentclass{aa}  

\usepackage{booktabs}
\usepackage{graphicx}
\usepackage{txfonts}
\def\kms {km\,s$^{-1}$ }

\begin{document}

   \title{First images of Antares photosphere  from spectropolarimetry}\titlerunning{Images of Antares photosphere}

    \author{{ Q.~Pilate}\inst{1},{ E.~Josselin}\inst{2},{ A.~Lèbre}\inst{2},{ A.~López Ariste}\inst{1},{ Ph.~Mathias}\inst{3}, {A.~Lavail}\inst{1}}

   \institute{
            Institut de Recherche en Astrophysique et Plan\'etologie,
            Universit\'e de Toulouse, UPS, CNRS, IRAP/UMR 5277,
            14 avenue Edouard Belin, F-31400, Toulouse, France  
            \and LUPM, UMR5299, CNRS and Université Montpellier, Place E.Bataillon, 34090 Montpellier, France \and IRAP, Universit\'e de Toulouse, CNRS, UPS, CNES, 57 avenue d'Azereix, 65000, Tarbes, France
             }

   \date{Received ...; accepted ...}

 
  \abstract
  {}
   { Antares is the closest red supergiant (RSG) to Earth. The discovery of linear polarization in the atomic lines of the star opens the path to produce direct images of the photosphere, hence probing the dynamics at the surface.    }
   {We analyze this linear polarization signals following the same scheme as has been previously done for the RSG Betelgeuse, and find that they are comparable in all its details. This allows us to use the same models for the analysis of these polarization signals in both stars.}
   {We found that as in Betelgeuse, the linear polarization signal of Antares is due to the depolarization of the continuum combined with brightness inhomogeneities. This allows us to produce images of the photosphere of star. We show that in Antares, convective cells can last several months and occupy roughly 30\% of the stellar radius for the largest ones.}
   {}


    \maketitle

%

\section{Introduction}

Red supergiants (RSGs) are among the brightest stars in the universe. Those cool evolved stars are able to enrich the interstellar medium through their stellar winds. However, the underlying mechanisms that trigger mass loss in RSGs are still poorly known.

Recently, a new technique to study RSGs was used for Betelgeuse. This technique consists in studying the linear polarization profile of the spectral lines of the star. This allowed \cite{Auriere16} to map bright spots at the surface, and later on, \cite{LopezAriste18} produced images of the photosphere and followed the time evolution of its surface convection. This new technique of imaging through spectropolarimetry requires the study of the mechanism responsible of the formation of linear polarization. In the case of Betelgeuse, the linear polarization signal was attributed to depolarization of the continuum by the atomic lines that absorb the light from a continuum polarized by Rayleigh scattering, and re-emit this light unpolarized \citep{Auriere16}. The depolarization produces an azimuthal symmetry, so that after integration over an uniform disk, any polarization signal would be canceled out. Thus, another mechanism that is able to break the spherical symmetry was invoked through the brightness of large convective cells, present at the surface of RSGs, as mentioned by \cite{Schwarzschild1975} and confirmed by observations \citep{Montarges16} and numerical simulations \citep{Freytag2002,Chiavassa22}. \\

In this article, we propose to apply this spectropolarimetric technique to the RSG Antares using spectropolarimetric data obtained at the Canada-France Hawaii Telescope (CFHT). In Sect.~\ref{Sect Antares} we describe Antares. In Sect.~\ref{2}, we study the linear polarization profile of Antares, from the least-squares deconvolution profiles and from the individual lines. We have a special look at the Na I D1 and D2 lines, that are known to depolarize the continuum and produce intrinsic polarization respectively. We also look at the signal in lines of Fe I, Ni I and Ca I that also depolarize the continuum. We then produce the first  images of the photosphere of Antares in Sect.~\ref{Sect2.2}. In Sect.~\ref{3} we conclude our work. 

\section{Antares}
\label{Sect Antares}
Antares ($\alpha$ Sco, M1.5 Iab), with Betelgeuse, is one of the prototypical RSG, and, at $\sim$ 170 pc \citep{2007A&A...474..653V}, 
maybe the nearest one. Indeed, both RSGs have similar spectral types and their position in the HR diagram are basically identical, 
suggesting similar temperature and surface gravity, but also similar masses \citep{Neuhauser22}, keeping in mind that 
the distance determination of RSGs may be quite uncertain \citep[see e.g.][]{2008AJ....135.1430H}. Based on the reinterpretation 
of antique observations, \cite{Neuhauser22} suggest that Betelgeuse may have been yellower about 2 millennia ago, 
suggesting a recent evolution toward the RSG and thus younger than Antares, which has been known as a red star for at least 3 
millennia.  However, \cite{2023MNRAS.526.2765S} propose that Betelgeuse could be in the core-carbon burning phase, and thus at 
the end of the RSG phase, before explosion as a supernova. At this stage, we thus just retain that Antares and Betelgeuse are similar in mass, temperature and gravity. 
Contrary to Betelgeuse, Antares has a wide binary companion ($\alpha$ Sco B; B2.5 V, \cite{1879Obs.....3...84J}) that is a hot main-sequence star located close to 2.73$"$ west of the RSG (Reimers et al. 2008). The orbit is seen nearly edge on, with a period of approximately 2700 years. Such binarity allows for a fairly precise determination of the mass-loss rate, $\dot{M}=7~10^{-7} ~M_\odot~\mathrm{yr}^{-1}$ \citep{1978A&A....70..227K}. 
Based on  ALMA and VLA observations, \cite{2020A&A...638A..65O} studied the upper atmosphere of both RSGs and found large-scale asymmetries, as well as signs of chromospheric activity. This should be linked to the supersonic turbulent motions found in Antares by 
\cite{Ohnaka17} who question a possible mechanism in addition to convection. In addition, \cite{Montarges17} determined the size of the convective cells at the surface of the star, which can reach up to 45$\%$ of the stellar radius.

\section{Spectropolarimetric data}
\label{2}

\subsection{Observations and data reduction}
Full Stokes observations of Antares were collected from February 2022 to July 2023 with ESPaDOnS \citep{2006ASPC..358..362D}, 
mounted at the Canada-France-Hawaii Telescope (CFHT), thus spanning 1.5 years. Each observation consisted in 4 sub-exposures 
of 2 sec for each Stokes parameter (Q, U and V) with 4 to 6 Q and U (linear polarization) and up to 23 V (circular polarization) 
observations at each date (see Table \ref{log of obs}), ensuring spectra with a signal-to noise ratio $\gtrsim 10^3$ per velocity bin 
as the signal in V is expected to be much weaker in the case of RSGs. The observations were unevenly spaced, in order to probe short-term (day to week) 
to long-term (month to year) variations. The spectra were extracted with Libre-ESpRIT \citep{Donati97}. The extracted spectra are the normalized intensity, and Stokes Q, U, V as a function of wavelength. Libre-ESpRIT also provides the Null spectra of the polarization signals. The null spectrum is, in principle, featureless and thus serves as a diagnostic tool for detecting spurious contributions in the polarized spectra. Additional information about the observation and data reduction methods can be found in \cite{Donati97}. 
A summary of the observation dates, with the corresponding polarimetric sequences is given in table~\ref{log of obs}.

\begin{table}[]
    \centering
    \begin{tabular}{lcc}
        \toprule
        \midrule
        Date & Sequence & $B_l$ (G) \\ 
        \midrule
        20 February 2022 & 2Q, 2U, 9V & $+1.27 \pm 0.3$ \\
        22 February 2022 & 4Q, 4U, 16V & $+1.14 \pm 0.3$ \\
        07 July 2022     & 4Q, 3U, 12V & $+1.87 \pm 0.3$ \\
        11 July 2022     & 4Q, 4U, 11V & $+1.56 \pm 0.3$  \\
        14 July 2022     & 4Q, 4U, 11V & $+2.10 \pm 0.3$ \\
        10 February 2023 & 5Q, 6U, 11V & $-1.99 \pm 0.3$ \\
        27 May 2023      & 2Q,4U, 22V & $-2.10 \pm 0.3$ \\
        24 June 2023     & 2Q, 2U, 20V & $-1.39 \pm 0.3$ \\
        27 June 2023     & 4Q, 3U, 21V & $-0.72 \pm 0.3$\\
        08 July 2023     & 2Q, 4U, 23V & $-1.28 \pm 0.3$\\
        \bottomrule
    \end{tabular}
    \caption{Log of observation of Antares with ESPaDOnS.}
    \label{log of obs}
\end{table}

\subsection{Mean profiles}

The study of the polarization in the individual lines of RSGs is often difficult, since the amplitude of linear polarization is roughly $10^{-4}$ times the continuum intensity and often close to noise level. In order to increase the signal to noise ratio, \cite{Donati97} proposed to sum many individual spectral lines, through a technique called Least-Squares Deconvolution (LSD). By assuming that for a given Stokes parameter, $Q$ or $U$ for the linear polarization, and $V$ for the circular polarization, the same signal up to a scale factor is present in every line, summing up will increase the signal-to-noise ratio. This leads to LSD profiles of Stokes $I$ for the intensity and Stokes $Q$, $U$ and $V$ for polarization. \\

\begin{figure}[!h]
    \centering
    \includegraphics[width=0.5\textwidth]{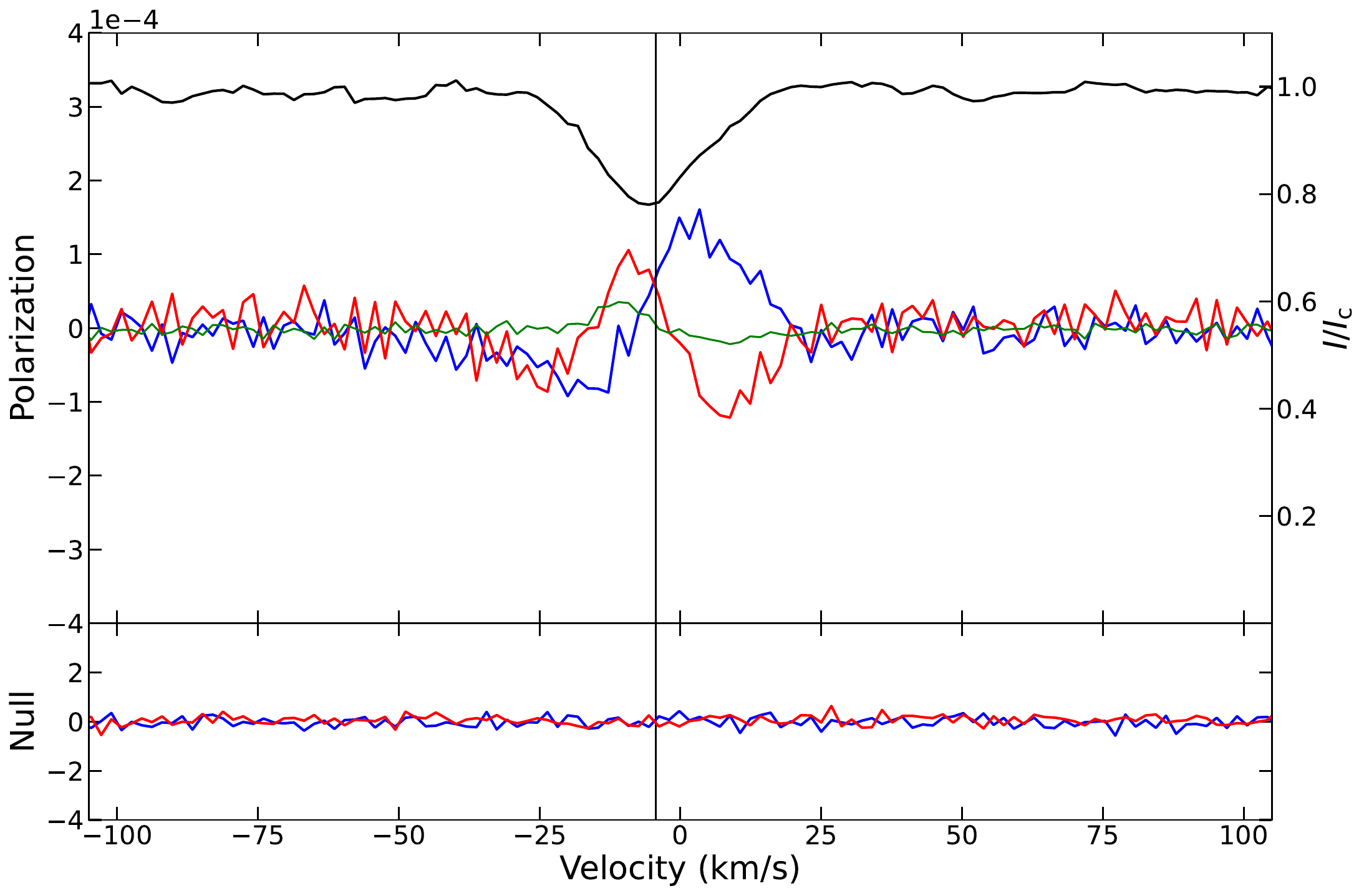}
    \caption{Top panel: LSD profile of Antares the 7th of July 2022. Stokes $U$ is in red, Stokes $Q$ in blue, Stokes $V$ in green and Stokes $I$ in black. The vertical black line represents the center of the line. Bottom panel: null profiles of Stokes $Q$ in blue and $U$ in red. The vertical black line depicts the velocity of the star at $-4.3~\mathrm{km.s^{-1}}$.}
    \label{LSD Antares juillet 2023}
\end{figure}

Since Betelgeuse and Antares have similar $\mathrm{T_{eff}}$ and $\log g$, we used the same mask as the one described by \cite{Auriere16}, from the Vienna Atomic Line Database \citep{Ryabchikova2015}, with an effective temperature of 3750~K, a surface gravity $\log g =0$ and a microturbulence of 4 $\mathrm{km.s^{-1}}$. The mask contains about 15 000 lines, with central depth deeper than 40$\%$ of the continuum intensity. The LSD profiles for the observations of the 7th of July 2022 is presented in Fig.~\ref{LSD Antares juillet 2023} (upper panel). The LSD profile of Stokes $Q$ is represented in blue, $U$ in red, $V$ in green and Stokes $I$ in black. The vertical black line depicts the velocity of the star reported by \cite{Pugh13} for reference, at -4.3 \kms. We see a clear linear polarization signal between -30\kms and 20\kms. 
The amplitude of linear polarization reaches up to $2 \times10^{-4}$ the continuum intensity, while Stokes $V$ only reaches $3 \times 10^{-5}$. The null profiles of both Stokes $Q$ and $U$ is shown in the bottom panel, with the same color code. The null profile of Stokes $V$ is not represented to avoid burdening the figure but reaches an amplitude up to $1 \times 10^{-5}$. \\

\subsection{Circular polarization}
The profiles have been normalised with a depth of 0.69, an equivalent Land\'e factor of 1.04 and an equivalent wavelength of 850 nm. 
Circular ((Stokes $V$, green line in Fig.~\ref{LSD Antares juillet 2023})) polarization is extracted with the least-squares deconvolution (LSD) method \citep{Donati97} and has been detected at each date. 
Despite the strong linear polarization signal, no significant cross-talk has been detected. 
Each Stokes $V$ profiles exhibit the typical Zeeman shape, represented as the green line in figure~\ref{ALL_LSD}. For clarity, the Stokes $V$ profiles have been shifted down and their amplitudes have been multiplied by a factor 5 to better see the profiles. The detection of Stokes $V$ allowed us to compute the mean longitudinal magnetic field $B_l$, which corresponds to the line-of-sight component of the magnetic field of the star, integrated over the stellar disk. We computed $B_l$ using the first-order moment of the Stokes $V$ Zeeman profile \citep[see e.g][]{Donati97,Wade2000}. We normalized the LSD profile using an average depth of 0.69, a mean Landé factor $g_\mathrm{eff}=1.04$ and a mean wavelength of 850 nm. The corresponding longitudinal fields are given in Table \ref{log of obs}.

The intensity reaches $< |B_l| > \, \approx 2$~G, comparable to the one detected toward Betelgeuse \citep{Mathias18}. With an uncertainty 
of $\sim$0.3~G, no significant variations on timescales $\lesssim$ week are seen. A polarity reversal occurred around the end of 
2022, suggesting a timescale of variation of $\sim$ 2 years.

The generation of a magnetic field in RSGs, which are very very slow rotators and thus have very high Rossby numbers 
is puzzling. However  \citep{2024ApJ...974..311A} have shown that a genuine dynamo can occur 
in low-gravity, fully convective stars. The ridges of the magnetic field are then aligned with the edges of the convective cells. 
Furthermore, their simulations show reversals of the field on timescales of a couple of years, consistent withe reversal observed for Antares. 

\subsection{Interpretation of linear polarization}
\label{Sect.depol}

To correctly interpret the origin of those linear polarization signals it is worth studying the polarization signal in the individual lines in the spectrum of Antares. Several mechanisms can produce linear polarization, while Stokes $V$ is almost uniquely related to the magnetic field and is routinely used to detect and map stellar magnetic fields \citep[e.g. ][]{Donati09}. Linear polarization due to Zeeman effect is roughly one order of magnitude smaller than Stokes $V$, but in Antares, the linear polarization amplitude is ten times higher than the Stokes $V$ signal (green line in Fig.~\ref{LSD Antares juillet 2023}), thus excluding any Zeeman origin of the linear polarization signal. Other mechanisms known to produce linear polarization in individual lines are found in the second solar spectrum \citep{Stenflo97}. One is the depolarization of the continuum by the atomic lines. In the Sun, roughly 90$\%$ of the lines depolarize the continuum. In the case of Betelgeuse, this mechanism was found to explain the origin of all linear polarization signals in atomic lines \citep{Auriere16}. Another mechanism is in the emission of polarized light, what we refer to as intrinsic polarization. While most of the lines are able to depolarize the continuum, only a small fraction of them can produce intrinsic polarization. This mechanism dominates the linear polarization of atomic lines in the spectrum of the Mira star $\chi$~Cyg \citep{LopezAriste19}. But, because of the similarities with Betelgeuse, we explore the continuum depolarization as the primary mechanism for the observed signals in Antares following \cite{Auriere16}.

\begin{figure}[!h]
    \centering
    \includegraphics[width=0.5\textwidth]{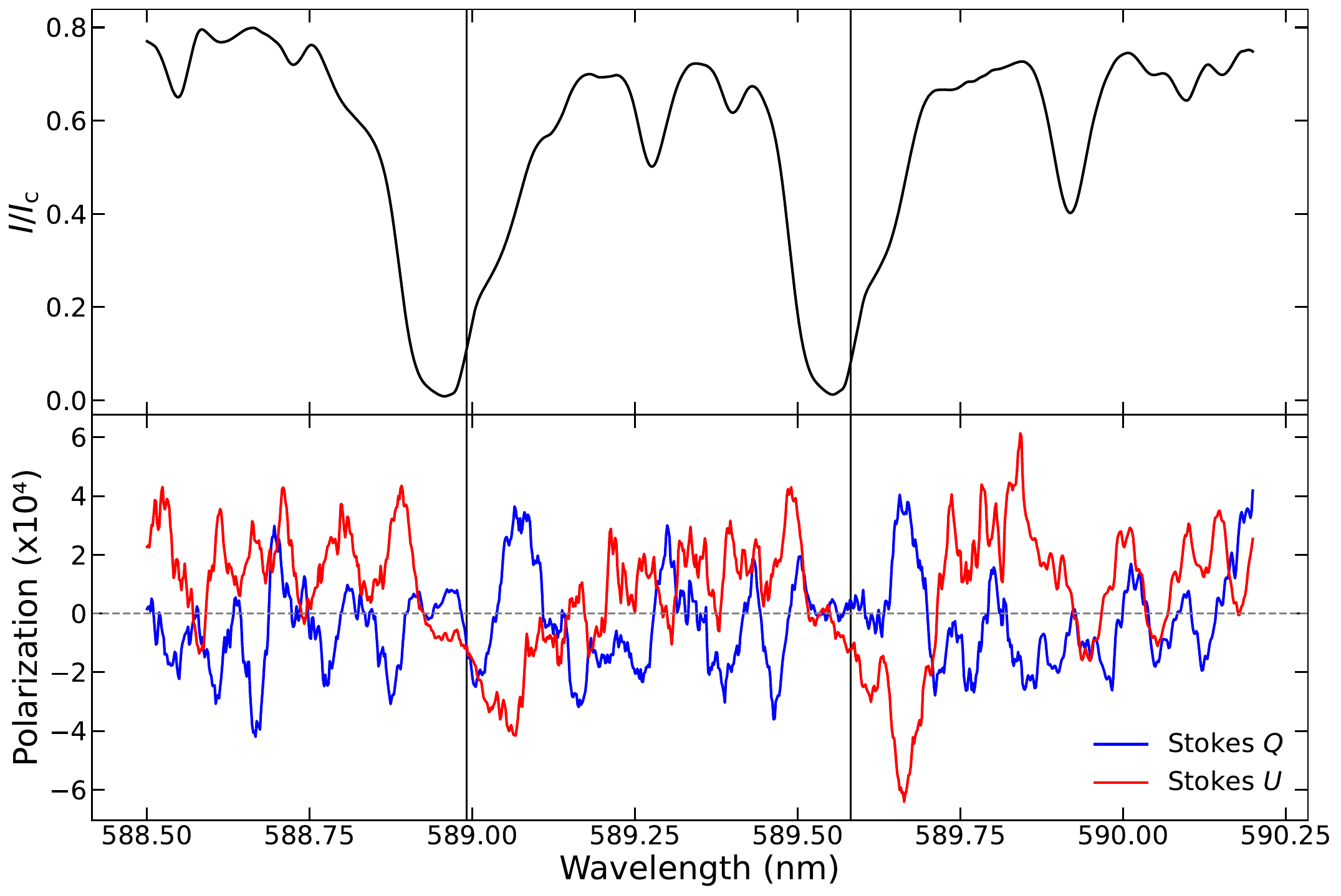}
    \caption{Polarized spectrum of Antares around Na I D1 and D2 lines. The upper panel depicts the intensity profile. The lower panel shows the linear polarization signal, with Stokes $Q$ in blue and Stokes $U$ in red. The vertical lines are the center of the lines D1 and D2, corrected from the Doppler velocity of -4.3 $\mathrm{km.s^{-1}}$.}
    \label{Sodium}
\end{figure}

One way to distinguish signals due to continuum polarization from intrinsic polarization is to look at the Na I lines D1 and D2. D1 cannot produce intrinsic polarization while D2 can \citep{Stenflo97,Landi04}. In Betelgeuse, D1 and D2 clearly exhibit similar amplitude in the linear polarization signals. Hence \cite{Auriere16} attributed this linear polarization signal to depolarization of the continuum. If we observe a polarization signal in D1 similar to that of D2 in the lines of Antares, the situation would be analogous to Betelgeuse. Figure.~\ref{Sodium} depicts the spectrum of Antares around the Na I doublet. The upper panel shows the intensity profile while the lower panel shows Stokes $U$ in red and Stokes $Q$ in blue. The amplitude of linear polarization in the individual lines being small, we added up the observations between the 7th of July 2022 and the 14th of July 2022. As will be shown in Sect.~\ref{Sect2.2}, the LSD profiles of Stokes $Q$ and $U$ do not significantly change within one week, allowing us to sum the profiles. The whole set consists on the average of thirteen $Q$ and $U$ profiles (see log of observations in Table~\ref{log of obs}). Note that we computed the average of the polarization signal within a range of $10^{-2}$ nm. Indeed, at those wavelength and with the spatial resolution of ESPaDOnS being R=68000, $\Delta\lambda=\lambda/\mathrm{R} \sim 10^{-2}$ nm, which allowed us to better see the profiles. Both D1 and D2 seem to produce a similar polarization profile, which is above the noise level. The two profiles show a positive lobe in $U$ and a negative one in $Q$ in the blue wing of the line, and vice versa in the red wing of the line, altogether with similar amplitudes. The Na~I D1 and D2 lines have been widely studied \citep[e.g,][and references therein]{Belluzzi15}, and the quantitative description of those lines is particularly hard, we thus restrict our analysis to the overall shape and amplitude of the Na~I lines. From this similarity of profile shapes and amplitudes, we attribute the linear polarization in Antares to depolarization of the continuum. \\

\begin{figure}[!h]
    \centering
    \includegraphics[width=0.5\textwidth]{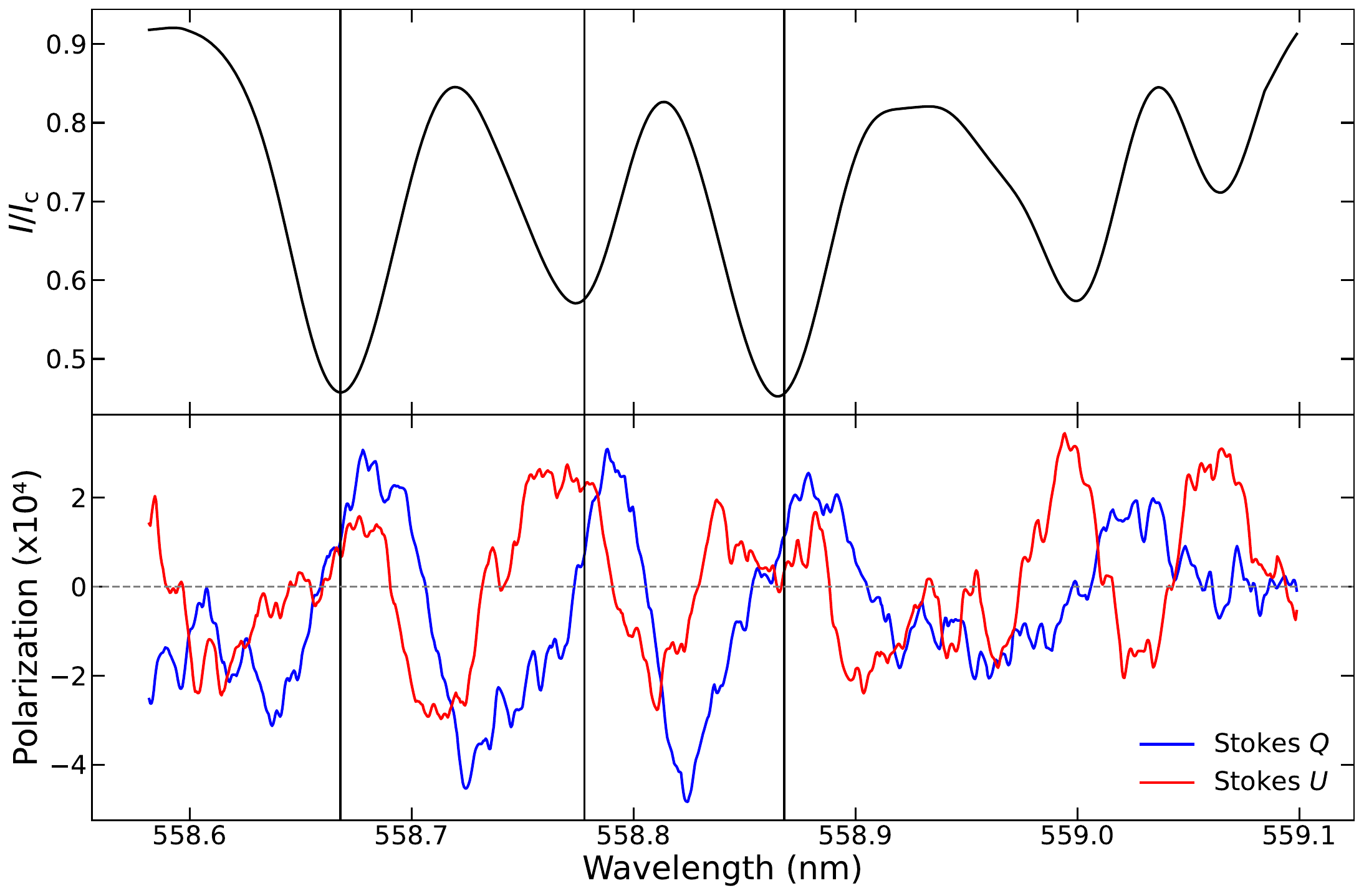}
    \caption{Intensity spectra around the Fe I, Ni I and Ca I lines (upper panel) and linear polarization spectrum in the lower panel, with Stokes $Q$ in blue and Stokes $U$ in red. The vertical lines represent the wavelength of the lines corrected with a Doppler shift of $-4.3~\mathrm{km.s^{-1}}$.}
    \label{Fe1Ni1Ca1}
\end{figure}

To ensure that the same linear polarization signal is present in the individual lines in Antares, we follow \cite{Auriere16} to look at the polarization signal in lines that are known to depolarize the continuum in the Sun \citep{Gandorfer2000}, namely Fe~I, Ni~I and Ca~I at 558.7 nm, 558.8 nm and 558.9 nm respectively. Once again, in Betelgeuse those lines exhibit a similar shape to the one of the LSD profile. Figure~\ref{Fe1Ni1Ca1} shows the intensity profile around those lines (upper panel) and the linear polarization in the lower panel, altogether with the center of the lines, Doppler shifted from their rest positions. As from D1 and D2 lines, we computed the average of linear polarization per bin of $10^{-2}$ nm, to better see the signal. Within the confidence limits given by the signal-to-noise ratios, the detected signals are similar between them and also similar to the signal extracted by the LSD procedure in Fig.~\ref{LSD Antares juillet 2023}, which justifies a posteriori, the choice of adding up the lines. Those three lines being known to depolarize the continuum and since the shape of the LSD profile and the shape of those individual lines are the same, we conclude that the linear polarization signal present in Antares is due to the depolarization of the continuum, and that we can add up the lines, independently of their physical properties.

\section{Imagery of the stellar photosphere}
\label{Sect2.2}

\subsection{LSD profiles and evolution}

\begin{figure*}[!h]
    \centering
    \includegraphics[width=\textwidth]{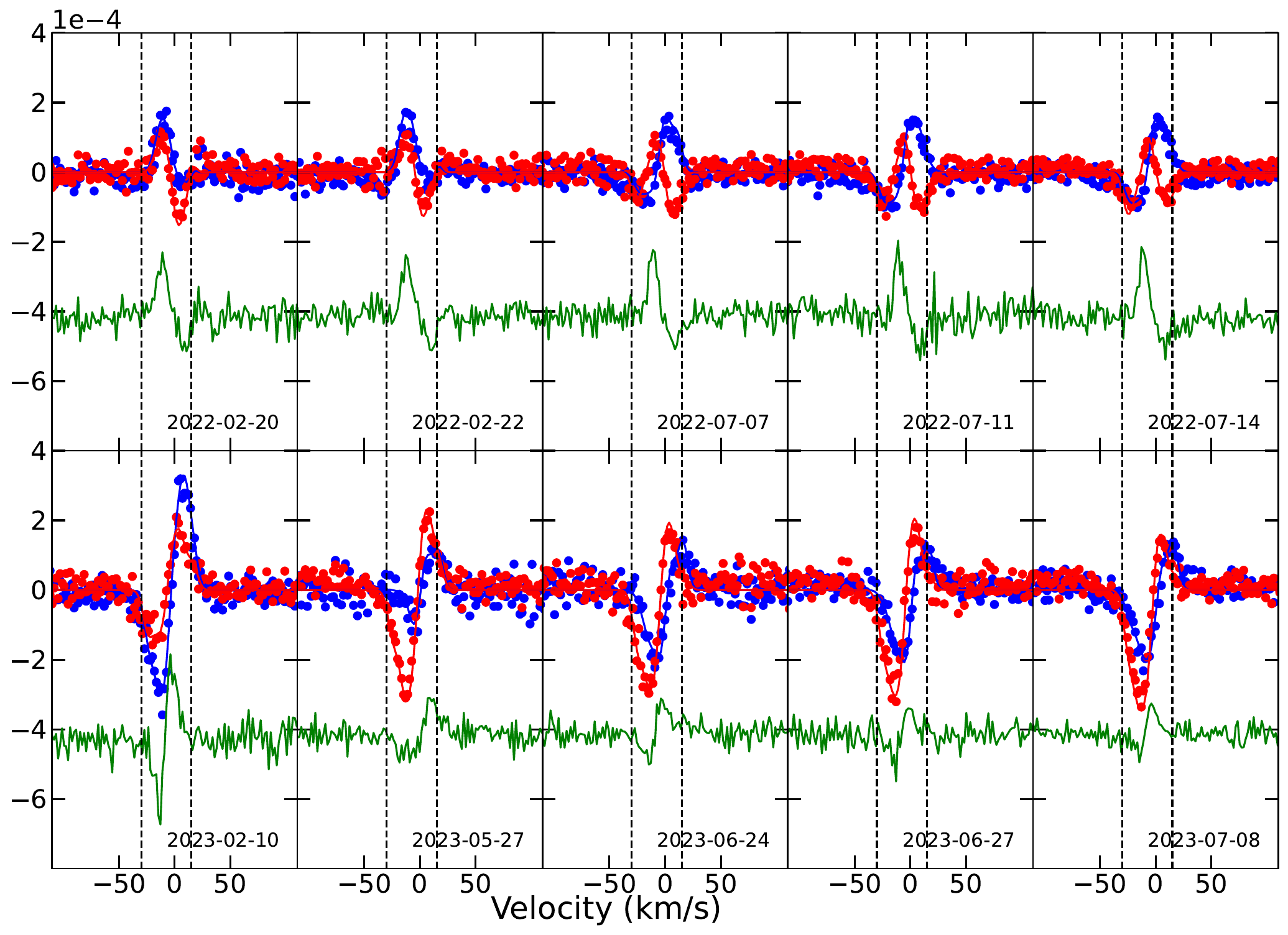}
    \caption{LSD profiles of Stokes $Q$ in blue, $U$ in red. The red and blue lines represent the brightness distribution that better fits the observed profile. The dates of observation are given in the bottom right of the figure. Stokes $V$ is represented in green and has been shifted down and its amplitude has been multiplied by 5 to better see the profiles. The vertical dashed lines represent from left to right, the velocity $v_0$ and the velocity of the star $v_*$ adopted in our model.}
    \label{ALL_LSD}
\end{figure*}

Figure~\ref{ALL_LSD} shows the LSD profiles of Antares from February 2022 to July 2023. Once again, the red and blue dots represent Stokes $U$ and $Q$ respectively, and the green line being Stokes $V$, shifted down and its amplitude has been multiplied by 5 for clarity. The log of observations for the different dates in presented in Table~\ref{log of obs}. The solid lines are the fit of the profiles, that we describe in next sub-section. The similarities of Antares with Betelgeuse continue: the LSD profiles seem to reach an amplitude between $2-4\times10^{-4}$ the continuum intensity, which is roughly the same order of magnitude as in Betelgeuse. The LSD profiles are varying within a span of a week to a month, which is clear between the 27th of May 2023 and the 24th of June 2023, where the amplitude of Stokes $Q$ (blue line) is increasing in the blue wing. We also pinpoint that the shape of the LSD profiles between the 7th of July and the 14th of July 2022 does not change significantly. This justifies our addition of lines in Sect.~\ref{Sect.depol}, since the same polarization signal is present in every observation. We did not represent the intensity profile in Fig.~\ref{ALL_LSD} to avoid burdening the figure. However, it is worth to compare it to the one of Betelgeuse and other RSGs. In Fig.~\ref{LSD Antares juillet 2023}, the intensity profile is narrow compared to the polarization profile. Especially, in the blue wing of the profile, around -25 \kms, there is a strong polarization signal, which falls in the continuum of the line. This behavior is present in every LSD profiles, but it is also present in Betelgeuse and other RSGs, such as $\mu$~Cep or CE~Tau and was attributed to velocity gradients, which narrow the intensity profile after disk integration \citep{LopezAriste25}.

\subsection{Images of the photosphere}

We found that the linear polarization signal in Antares is the same in every lines. Furthermore, the polarization signal was attributed to the depolarization of the continuum by the atomic lines present in the photosphere. We recall that the depolarization of the continuum produces an azimuthal symmetry, hence after integration over the stellar disk, any polarization signal would be canceled out. Thus, we have to introduce a mechanism that breaks the symmetry of the disk. In RSGs, the large convective cells theorized by \cite{Schwarzschild1975} were used by \cite{Auriere16} to explain the breaking of symmetry over the disk. This allowed \cite{LopezAriste18} to produce images of Betelgeuse. All the similarities pointed out above between Antares and Betelgeuse lead us to apply the same models here to produce images of the photospheric brightness distribution of Antares. Before going any further, we recall the imaging technique. The technique involves finding the brightness distribution that better fits the observed LSD profiles. The imaging technique also requires the introduction of two velocities: first, the maximum velocity of the rising plasma. Since we assume that most of the photons that we receive come from hot rising plasma, we chose an ad hoc velocity $v_0$ that is the maximum velocity of the rising plasma. This velocity is chosen by assuming that no polarization signal can be blueshifted compared to this velocity. We choose a value of -45~\kms, consistent with the work of \cite{Ohnaka17}, who found similar velocity amplitude in the outer layers of Antares. The second velocity is the velocity of the star $v_\star$. It is tempting to chose the classical value of -4.3 \kms \citep{Pugh13}, which represents the center of the line, and which is represented in Fig.~\ref{LSD Antares juillet 2023}. However, in RSGs, the shape of the intensity line is not symmetric, as pointed out by \cite{Josselin07}. In RSGs such as Betelgeuse, the line exhibits the classical C-shape bisector, associated with convection. The same behavior is also present in Antares. Therefore, in RSGs, the velocity of the star lies in the red wing of the line \citep{LopezAriste18}. Hence, we chose $v_\star=15$ \kms, and altogether, in the heliocentric reference frame of the LSD profiles, $v_0$ is at -30 \kms. As discussed in \cite{LopezAriste18}, each point in the disk was supposed to emit a gaussian profile, with an FWHM of 10 \kms, to take into account instrumental width, thermal broadening and macroturbulence. \\

\begin{figure}[!h]
    \centering
    \includegraphics[width=0.5\textwidth]{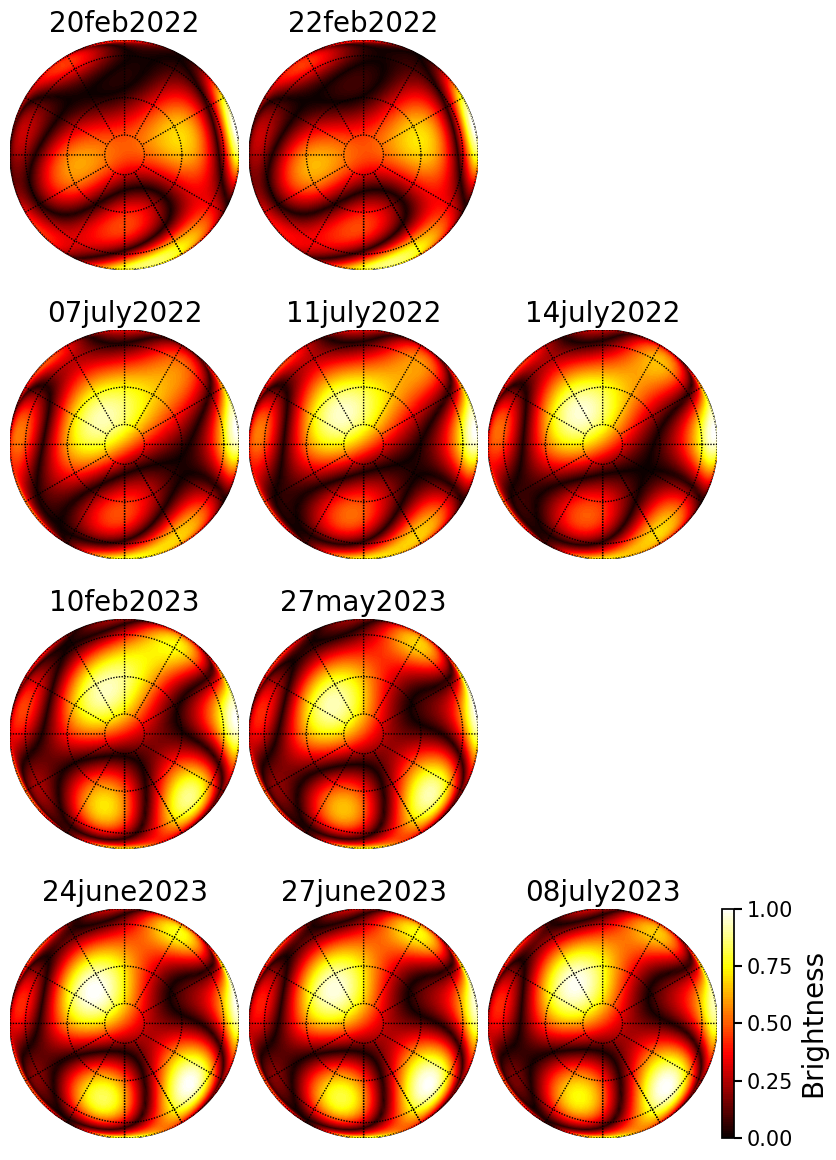}
    \caption{Images of Antares. North is up, East is left. In each images, the brightness is normalized.}
    \label{Images Antares}
\end{figure}

Figure~\ref{Images Antares} shows the images of the photosphere of Antares, from February 2022 to July 2023. The associated Stokes profiles of each images are represented with the solid lines in Fig.~\ref{ALL_LSD}, where it is clear that the reconstructed images fit well the data points. In every images, the brightness was normalized. The brightness distribution was modeled using spherical harmonics up to $l_\mathrm{max}=5$ \citep{LopezAriste18}. As in \cite{LopezAriste22}, the brightness is correlated with velocity, 75$\% $of the plasma is moving upward while 25$\%$ is falling to the star. The higher the brightness, the faster the plasma is moving upward. Therefore, the dark areas in the images represent the zones where the plasma is falling toward Antares. Looking at the images, the brightness distribution changes with a span of a several months. The convective cell near disk center is consistent with the $v_0$ that we chose, since the peak in linear polarization is close to this velocity. The central convective cell is present for at least one year, which is consistent with numerical simulations \citep{Freytag2002}, where the large convective cells have a typical lifetime around two years. In February 2023, another large convective cell appears in the Southwestern hemisphere. This is also consistent with the LSD profile at the same date, where a strong polarization signal appears in Stokes $Q$. Since the peak is located closer to our ad hoc velocity of the star, is appears closer to the stellar limb. Overall, our images exhibit between one convective cell (July 2022) and three (June 2023), once again, consistent with numerical simulations and theory \citep{Schwarzschild1975,Freytag2002}. The size of the convective cells reach up to 30$\%$ of the stellar radius in our images. This size was estimated by computing the area of the largest convective cell on the image obtained the 8th of July 2023, located in northeastern hemisphere of the star.  \\

And as with the analogous images in Betelgeuse, a word of caution is required: one must not over interpret those images. Linear polarization carries a $180^{\circ}$ ambiguity, thus every images can be rotated by $180^{\circ}$ and the brightness distribution will still fit the observed profile. The combination of both spectropolarimetry and interferometry can allow us to avoid any ambiguity. Despite those ambiguities, the overall position of the convective cell should be correct, within $180^{\circ}$ ambiguity. To avoid any ambiguities in the reconstruction of the images, for a given day, we used the brightness distribution found in the previous observation to fit the observed linear polarization profile. This ensures a continuity between the images. Such technique was also used by \cite{LopezAriste18} to follow the surface convection of Betelgeuse. 

\section{Conclusion}
\label{3}

The red supergiant Antares exhibits a strong linear polarization signal. This signal has an amplitude ten times higher than the Stokes $V$ signal, excluding any Zeeman origin of the linear polarization. Looking at the Na I D1 and D2 lines, we showed that both lines show a clear polarization signal. The D1 line can only depolarize the continuum, while D2 can produce intrinsic polarization. By studying the three lines Fe I, Ni I and Ca I, that are known to depolarize the sun's continuum \citep{Gandorfer2000}, we showed that a clear linear polarization appears in those lines, and that the shape of the profile is the same as the one observed in the LSD profile. Hence, we were able to affirm that the linear polarization that we observed in Antares is due to the depolarization of the continuum by the atomic lines present in the photosphere. All these facts from Antares are perfectly comparable to those of Betelgeuse \citep{LopezAriste18} and we were able to apply the same techniques and produce the first images of the brightness distribution in the photosphere of Antares at different epochs and track its surface convection. The velocity span observed in the linear polarization seems to be comparable to the velocity amplitude given by \cite{Ohnaka17}. The size and the number of cells are consistent with the work of \cite{Schwarzschild1975}. The technique of spectropolarimetry imaging was successfully applied to Antares, and can be extended to other RSGs, as long as their lines depolarize the continuum. This technique has to be combined with other techniques of imaging, such as interferometry, to avoid any ambiguity in the reconstructed images.

\begin{acknowledgements}
    This work was supported by the "Programme National de Physique Stellaire" (PNPS) of CNRS/INSU co-funded by CEA and CNES.
    We acknowledge support from the French National Research Agency (ANR)
    funded project PEPPER (ANR-20-CE31-0002).

    \end{acknowledgements}
    
    \bibliographystyle{aa}
    
    \bibliography{main}

\begin{thebibliography}{31}
\expandafter\ifx\csname natexlab\endcsname\relax\def\natexlab#1{#1}\fi

\bibitem[{{Amard} {et~al.}(2024){Amard}, {Brun}, \& {Palacios}}]{2024ApJ...974..311A}
{Amard}, L., {Brun}, A.~S., \& {Palacios}, A. 2024, \apj, 974, 311

\bibitem[{{Auri{\`e}re} {et~al.}(2016){Auri{\`e}re}, {L{\'o}pez Ariste}, {Mathias}, {L{\`e}bre}, {Josselin}, {Montarg{\`e}s}, {Petit}, {Chiavassa}, {Paletou}, {Fabas}, {Konstantinova-Antova}, {Donati}, {Grunhut}, {Wade}, {Herpin}, {Kervella}, {Perrin}, \& {Tessore}}]{Auriere16}
{Auri{\`e}re}, M., {L{\'o}pez Ariste}, A., {Mathias}, P., {et~al.} 2016, \aap, 591, A119

\bibitem[{{Belluzzi} {et~al.}(2015){Belluzzi}, {Trujillo Bueno}, \& {Landi Degl'Innocenti}}]{Belluzzi15}
{Belluzzi}, L., {Trujillo Bueno}, J., \& {Landi Degl'Innocenti}, E. 2015, \apj, 814, 116

\bibitem[{{Chiavassa} {et~al.}(2022){Chiavassa}, {Kudritzki}, {Davies}, {Freytag}, \& {de Mink}}]{Chiavassa22}
{Chiavassa}, A., {Kudritzki}, R., {Davies}, B., {Freytag}, B., \& {de Mink}, S.~E. 2022, \aap, 661, L1

\bibitem[{{Donati} {et~al.}(2006){Donati}, {Catala}, {Landstreet}, \& {Petit}}]{2006ASPC..358..362D}
{Donati}, J.~F., {Catala}, C., {Landstreet}, J.~D., \& {Petit}, P. 2006, in Astronomical Society of the Pacific Conference Series, Vol. 358, Solar Polarization 4, ed. R.~{Casini} \& B.~W. {Lites}, 362

\bibitem[{{Donati} \& {Landstreet}(2009)}]{Donati09}
{Donati}, J.~F. \& {Landstreet}, J.~D. 2009, \araa, 47, 333

\bibitem[{{Donati} {et~al.}(1997){Donati}, {Semel}, {Carter}, {Rees}, \& {Collier Cameron}}]{Donati97}
{Donati}, J.~F., {Semel}, M., {Carter}, B.~D., {Rees}, D.~E., \& {Collier Cameron}, A. 1997, \mnras, 291, 658

\bibitem[{{Freytag} {et~al.}(2002){Freytag}, {Steffen}, \& {Dorch}}]{Freytag2002}
{Freytag}, B., {Steffen}, M., \& {Dorch}, B. 2002, Astronomische Nachrichten, 323, 213

\bibitem[{{Gandorfer}(2000)}]{Gandorfer2000}
{Gandorfer}, A. 2000, {The Second Solar Spectrum: A high spectral resolution polarimetric survey of scattering polarization at the solar limb in graphical representation. Volume I: 4625 {\r{A}} to 6995 {\r{A}}}

\bibitem[{{Harper} {et~al.}(2008){Harper}, {Brown}, \& {Guinan}}]{2008AJ....135.1430H}
{Harper}, G.~M., {Brown}, A., \& {Guinan}, E.~F. 2008, \aj, 135, 1430

\bibitem[{{Johnson}(1879)}]{1879Obs.....3...84J}
{Johnson}, S.~J. 1879, The Observatory, 3, 84

\bibitem[{{Josselin} \& {Plez}(2007)}]{Josselin07}
{Josselin}, E. \& {Plez}, B. 2007, \aap, 469, 671

\bibitem[{{Kudritzki} \& {Reimers}(1978)}]{1978A&A....70..227K}
{Kudritzki}, R.~P. \& {Reimers}, D. 1978, \aap, 70, 227

\bibitem[{{Landi Degl'Innocenti} \& {Landolfi}(2004)}]{Landi04}
{Landi Degl'Innocenti}, E. \& {Landolfi}, M. 2004, {Polarization in Spectral Lines}, Vol. 307

\bibitem[{{L{\'o}pez Ariste} {et~al.}(2022){L{\'o}pez Ariste}, {Georgiev}, {Mathias}, {L{\`e}bre}, {Wavasseur}, {Josselin}, {Konstantinova-Antova}, \& {Roudier}}]{LopezAriste22}
{L{\'o}pez Ariste}, A., {Georgiev}, S., {Mathias}, P., {et~al.} 2022, \aap, 661, A91

\bibitem[{{L{\'o}pez Ariste} {et~al.}(2018){L{\'o}pez Ariste}, {Mathias}, {Tessore}, {L{\`e}bre}, {Auri{\`e}re}, {Petit}, {Ikhenache}, {Josselin}, {Morin}, \& {Montarg{\`e}s}}]{LopezAriste18}
{L{\'o}pez Ariste}, A., {Mathias}, P., {Tessore}, B., {et~al.} 2018, \aap, 620, A199

\bibitem[{{L{\'o}pez Ariste} {et~al.}(2019){L{\'o}pez Ariste}, {Tessore}, {Carl{\'\i}n}, {Mathias}, {L{\`e}bre}, {Morin}, {Petit}, {Auri{\`e}re}, {Gillet}, \& {Herpin}}]{LopezAriste19}
{L{\'o}pez Ariste}, A., {Tessore}, B., {Carl{\'\i}n}, E.~S., {et~al.} 2019, \aap, 632, A30

\bibitem[{{López Ariste} {et~al.}(2025){López Ariste}, {Pilate}, {Lavail}, \& {Mathias}}]{LopezAriste25}
{López Ariste}, A., {Pilate}, Q., {Lavail}, A., \& {Mathias}, P. 2025, arXiv e-prints, arXiv:2503.18504

\bibitem[{{Mathias} {et~al.}(2018){Mathias}, {Auri{\`e}re}, {L{\'o}pez Ariste}, {Petit}, {Tessore}, {Josselin}, {L{\`e}bre}, {Morin}, {Wade}, {Herpin}, {Chiavassa}, {Montarg{\`e}s}, {Konstantinova-Antova}, {Kervella}, {Perrin}, {Donati}, \& {Grunhut}}]{Mathias18}
{Mathias}, P., {Auri{\`e}re}, M., {L{\'o}pez Ariste}, A., {et~al.} 2018, \aap, 615, A116

\bibitem[{{Montarg{\`e}s} {et~al.}(2017){Montarg{\`e}s}, {Chiavassa}, {Kervella}, {Ridgway}, {Perrin}, {Le Bouquin}, \& {Lacour}}]{Montarges17}
{Montarg{\`e}s}, M., {Chiavassa}, A., {Kervella}, P., {et~al.} 2017, \aap, 605, A108

\bibitem[{{Montarg{\`e}s} {et~al.}(2016){Montarg{\`e}s}, {Kervella}, {Perrin}, {Chiavassa}, {Le Bouquin}, {Auri{\`e}re}, {L{\'o}pez Ariste}, {Mathias}, {Ridgway}, {Lacour}, {Haubois}, \& {Berger}}]{Montarges16}
{Montarg{\`e}s}, M., {Kervella}, P., {Perrin}, G., {et~al.} 2016, \aap, 588, A130

\bibitem[{{Neuh{\"a}user} {et~al.}(2022){Neuh{\"a}user}, {Torres}, {Mugrauer}, {Neuh{\"a}user}, {Chapman}, {Luge}, \& {Cosci}}]{Neuhauser22}
{Neuh{\"a}user}, R., {Torres}, G., {Mugrauer}, M., {et~al.} 2022, \mnras, 516, 693

\bibitem[{{O'Gorman} {et~al.}(2020){O'Gorman}, {Harper}, {Ohnaka}, {Feeney-Johansson}, {Wilkeneit-Braun}, {Brown}, {Guinan}, {Lim}, {Richards}, {Ryde}, \& {Vlemmings}}]{2020A&A...638A..65O}
{O'Gorman}, E., {Harper}, G.~M., {Ohnaka}, K., {et~al.} 2020, \aap, 638, A65

\bibitem[{{Ohnaka} {et~al.}(2017){Ohnaka}, {Weigelt}, \& {Hofmann}}]{Ohnaka17}
{Ohnaka}, K., {Weigelt}, G., \& {Hofmann}, K.~H. 2017, \nat, 548, 310

\bibitem[{{Pugh} \& {Gray}(2013)}]{Pugh13}
{Pugh}, T. \& {Gray}, D.~F. 2013, \aj, 145, 38

\bibitem[{{Ryabchikova} {et~al.}(2015){Ryabchikova}, {Piskunov}, {Kurucz}, {Stempels}, {Heiter}, {Pakhomov}, \& {Barklem}}]{Ryabchikova2015}
{Ryabchikova}, T., {Piskunov}, N., {Kurucz}, R.~L., {et~al.} 2015, \physscr, 90, 054005

\bibitem[{{Saio} {et~al.}(2023){Saio}, {Nandal}, {Meynet}, \& {Ekstr{\"o}m}}]{2023MNRAS.526.2765S}
{Saio}, H., {Nandal}, D., {Meynet}, G., \& {Ekstr{\"o}m}, S. 2023, \mnras, 526, 2765

\bibitem[{{Schwarzschild}(1975)}]{Schwarzschild1975}
{Schwarzschild}, M. 1975, \apj, 195, 137

\bibitem[{{Stenflo} \& {Keller}(1997)}]{Stenflo97}
{Stenflo}, J.~O. \& {Keller}, C.~U. 1997, \aap, 321, 927

\bibitem[{{van Leeuwen}(2007)}]{2007A&A...474..653V}
{van Leeuwen}, F. 2007, \aap, 474, 653

\bibitem[{{Wade} {et~al.}(2000){Wade}, {Donati}, {Landstreet}, \& {Shorlin}}]{Wade2000}
{Wade}, G.~A., {Donati}, J.~F., {Landstreet}, J.~D., \& {Shorlin}, S.~L.~S. 2000, \mnras, 313, 851

\end{thebibliography}
\end{document}